\documentclass[aps,prl,superscriptaddress,twocolumn,floatfix,noeprint]{revtex4-1}

\usepackage[utf8]{inputenc}
\usepackage[T1]{fontenc}
\usepackage{mathtools}
\usepackage{dsfont}
\usepackage{bm}
\usepackage{graphicx}
\usepackage[breaklinks]{hyperref}
\usepackage[dvipsnames]{xcolor}
\usepackage{tikz}
\usepackage{url}
\usepackage{comment}
\usepackage[export]{adjustbox}
\usepackage{float}
\usepackage[caption = false]{subfig}

\newcommand{\bvec}[1]{\bm{#1}}

\usepackage{color}

\begin{document}

\allowdisplaybreaks

\title{Spin-fluctuation-induced pairing in twisted bilayer graphene}
\author{Ammon Fischer} 
\affiliation{Institute for Theoretical Solid State Physics,
RWTH Aachen University, and JARA Fundamentals
of Future Information Technology, 52062 Aachen, Germany}
\author{Lennart Klebl} 
\affiliation{Institute for Theory of Statistical Physics,
RWTH Aachen University, and JARA Fundamentals
of Future Information Technology, 52062 Aachen, Germany}
\author{Carsten Honerkamp}
\affiliation{Institute for Theoretical Solid State Physics,
RWTH Aachen University, and JARA Fundamentals
of Future Information Technology, 52062 Aachen, Germany}
\author{Dante M. Kennes}
\affiliation{Institute for Theory of Statistical Physics,
RWTH Aachen University, and JARA Fundamentals
of Future Information Technology, 52062 Aachen, Germany}
\affiliation{Max Planck Institute for the Structure and Dynamics of Matter, Center for Free Electron Laser Science, 22761 Hamburg, Germany}

\date{\today}

\begin{abstract}
 We investigate the interplay of magnetic fluctuations and Cooper pairing in twisted bilayer graphene from a purely microscopic model within a large-scale tight-binding approach resolving the \AA ngstr\"om scale. For local onsite repulsive interactions and using the random-phase approximation for spin fluctuations, we derive a microscopic effective pairing interaction that we use for self-consistent solutions of the Bogoliubov-de-Gennes equations of superconductivity. We study the predominant pairing types as function of interaction strength, temperature and band filling. For large regions of this parameter space, we find chiral $d$-wave pairing regimes, spontaneously breaking time-reversal symmetry, separated by magnetic instabilities at integer band fillings. Interestingly, the $d$-wave pairing is strongly concentrated in the AA regions of the moiré unit cell and exhibits phase windings of integer multiples of $2\pi$ around these superconducting islands, i.e. pinned vortices. The spontaneous circulating current creates a distinctive magnetic field pattern. This signature of the chiral pairing should be measurable by state-of-the-art experimental techniques. 
\end{abstract}

\maketitle

\emph{Introduction.---}
Twisted two-dimensional materials have become an extremely active research area, with twisted bilayer \cite{cao2018a,Yankowitz18,lu2019superconductors,stepanov2019interplay,cao2018b,lu2019superconductors,stepanov2019interplay,sharpe19,serlin19,Xu18,Liu18,Kennes2018d0,Li2010,Kerelsky18,xie2019spectroscopic,jiang2019charge,Choi19, PhysRevResearch.2.033062, PhysRevB.102.064501}, trilayer \cite{shimazaki19}
and double bilayer graphene \cite{huang18b,kerelsky19, shen2020correlated, burg2019correlated, liu2020tunable, cao2020tunable, he2020tunable}, transition metal dichalcogenides homo- and heterobilayers \cite{wang19,regan2020,tang2020,xian20} as well as other materials \cite{Xian18,Ni2019,kennes19} at the frontier of condensed matter research. These systems are fascinating due to high degree of band-structure and correlation engineering that can be achieved, putting unprecedented topological and exotic correlated states within experimental reach \cite{moiresim}. In fact, twisting can lead to strong alterations of the band structures, most notably by the formation of flat bands, that may enhance various interaction effects and hence lead to interesting and potentially novel, interaction-driven ground states \cite{moiresim}. From a theoretical point of view, the twisting-induced large moiré unit cell poses a formidable challenge for the description on the atomic scale as an excessive amount of degrees of freedoms have to be treated.

With regard to twisted bilayer graphene (TBG), which initially triggered this immense research activities on twisted 2D heterostructures, the main questions about the quantum many-body state at low temperatures still concerns the nature of the insulating and superconducting states found experimentally. Recently, we have addressed the question in the context of insulating states by microscopic random phase approximation (RPA) \cite{klebl2019inherited} and functional renormalization group (fRG) techniques \cite{klebl2020functional}, keeping the full unit cell containing many thousands atoms (depending on the twist angle) under microscopic scrutiny. This microscopic theory on the carbon-carbon bond scale predicts the leading electronic interaction-driven instabilities for a larger parameter range to be magnetically ordered  with strong spatial variations of the order parameter through the large moiré unit cell. However, in the same approach, superconducting pairing instabilities have not received much attention, beyond a re-scaled theory losing some of the connection to the microscopics \cite{PhysRevB.98.195101, wolf2019electrically, gonzalez2017electrically}. 



Thus, despite the enormous theoretical effort in the young history of this field, fully microscopic models that capture the full electronic spectrum of TBG whilst providing a mechanism for strong electron-electron interaction are still rare. In this Letter, we remedy this shortcoming by (i) using a full tight-binding approach for the $\pi$-bands of TBG, (ii) deriving an effective two-particle interaction vertex $\Gamma_2$ by using the RPA to include spin-fluctuation exchange between electrons to high orders in the bare couplings and (iii) using a mean-field decoupling of this effective interaction to analyze the nature and experimental signatures of the favored superconducting state in a fully microscopic model. 
By considering spin-fluctuations alone as the potential pairing glue we do neglect other forms of electronic two-particle scattering. 
When considering onsite bare electron-electron interactions only, this approximation  was recently shown to be justified comparing RPA and unbiased fRG techniques \cite{klebl2019inherited, klebl2020functional}, but of course including phonons would widen the range of possibilities beyond this study.  
From the derived electronically mediated pairing interaction we extract pairing symmetries and the spatial distribution of the superconducting order parameter on the carbon-carbon bond scale, with the advantage of this approach being that it allows to study both, magnetic instabilities and superconducting pairing as function of temperature and doping. Our work thus establishes a microscopic scenario of electronic correlation effects leading to the phase diagrams of TBG recently measured in experiments \cite{cao2018b, cao2018a, lu2019superconductors} and provides direct predictions to be tested.
As a function of the carrier density, between the insulating states at integer fillings, we find an attractive spin-fluctuation mediated pairing vertex that gives rise to chiral $d$-wave pairing regimes, spontaneously breaking time-reversal symmetry (TRS). The superconducting order parameter is strongly enhanced in the AA regions of the system and exhibits supercurrents within a vortex-antivortex structure as well as magnetic fields, which yield measurable signatures of the chiral pairing in experiment. This will help to distinguish the origin of superconductivity in TBG and settle the debate about whether its mechanism is electron- or phonon-driven \cite{stepanov2019interplay}. 

\paragraph{Methods.---}

For our fully microscopic theory we first follow Ref.~\onlinecite{PhysRevX.8.031087} to set up an atomistic tight-binding Hamiltonian for the $\pi$-band spectrum of magic-angle TBG ($\theta = 1.05^{\circ}$) keeping all $N=11908$ bands under consideration
\begin{equation}
H_0 = \sum_{\bvec{R}, \bvec{R'}} \sum_{i,j, \sigma} t(\bvec{R} + \bvec{r}_i - \bvec{R'} - \bvec{r}_j) c_{\bvec{R}, \bvec{r}_i, \sigma}^{\dagger} c_{\bvec{R'}, \bvec{r}_j, \sigma}^{\phantom \dagger}.
\label{H0}
\end{equation}
The hopping parameters decay exponentially on the carbon-carbon bond scale (see the Supplement Material \footnote{See Supplemental Material at
\url{http://link.aps.org/supplemental/10.1103/PhysRevB.103.L041103} for numerical details of the atomistic tight-binding model and RPA/MF calculations, including Ref. \cite{PhysRevX.8.031087, klebl2019inherited, berk1966effect, romer2012local, zhu2016bogoliubov, ozaki2007, cunningham1974, uchida2014,PhysRevB.48.17427, moon2013optical, dos2012continuum, sboychakov2015electronic}} for details) and are derived from first-principle calculations \cite{PhysRevX.8.031087}. When taking atomic relaxation effects between the layers into account, the spectrum contains four flat bands (two-fold spin degenerate) around charge neutrality, separated from the rest of the spectrum.
 
 We include interaction effects by a repulsive Hubbard term for electrons with opposite spin $\sigma$ residing on the same carbon site
\begin{equation}
H_{\text{int}} = \frac{1}{2}\sum_{\bvec{R},i, \sigma} U n_{\bvec{R}, \bvec{r}_i, \sigma} n_{\bvec{R}, \bvec{r}_i, \overline{\sigma}},
\label{hubbard}
\end{equation}
where $\bvec{R}$ labels the supercell vector and $\bvec{r}_i$ is restricted to the moir\'{e} unit cell. This onsite term is an idealization of the true long-ranged nature of the actual Coulomb interaction. Theoretical work for non-twisted systems \cite{TangStrain,SanchezdlP} indicates that the main instabilities towards insulating states are correctly captured by this idealization. The value for $U$ is well established within the cRPA approximation for  mono- and bilayer graphene \cite{Wehling2011, Roesner} and the effect of the non-local terms may be absorbed into an effective $U^{*}$ \cite{Schueler}.

To characterize the potential ground state of the interacting system, a two-step protocol is employed. First, we study spin fluctuations and associated magnetic ordering of TBG by using the random-phase approximation (RPA) for the magnetic susceptibility $\hat{\chi}(\bvec{q}, \nu)$. Here, we exploit the methodology proposed in Ref.~\onlinecite{klebl2019inherited} that captures spin-fluctuations on the carbon-carbon bond scale with emphasis on the static, long-wavelength limit ($\bvec q,\nu \to 0$) on the moir\'e scale.
The latter limit proves to contain the relevant physics when starting with local repulsive interaction. The RPA susceptibility predicts spin correlations at length scales intermediate to the c-c bond scale and moir\'e length scale, thus being described by orderings at $\bvec q=0$. The system shows the same order in all moir\'{e} unit cells with variable correlations present on the C-C bond scale.


In a second step, we deduce an effective singlet pairing vertex $\hat{\Gamma}_2(\bvec{q},\nu)$ from transverse and longitudinal spin-fluctuations \cite{berk1966effect, romer2012local}
\begin{equation}
\hat{\Gamma}_2(\bvec{q}, \nu) = \hat{U}  - \frac{U^2\hat{\chi}_0(\bvec{q}, \nu)}{1+U\hat{\chi}_0(\bvec{q}, \nu)}  + \frac{U^3\hat{\chi}_0^2(\bvec{q}, \nu)}{1-U^2\hat{\chi}_0^2(\bvec{q}, \nu)}, 
\label{eq_veff}
\end{equation}

and analyze the vertex by using a mean-field decoupling to extract pairing symmetries and spatial distribution of the superconducting order parameter in the moir\'{e} unit cell. Again, we neglect the momentum dependence of the interaction vertex and thus focus on the pairing structure on the c-c bonds within the moiré unit cell.


\begin{figure}
    \includegraphics[width = 0.48\textwidth]{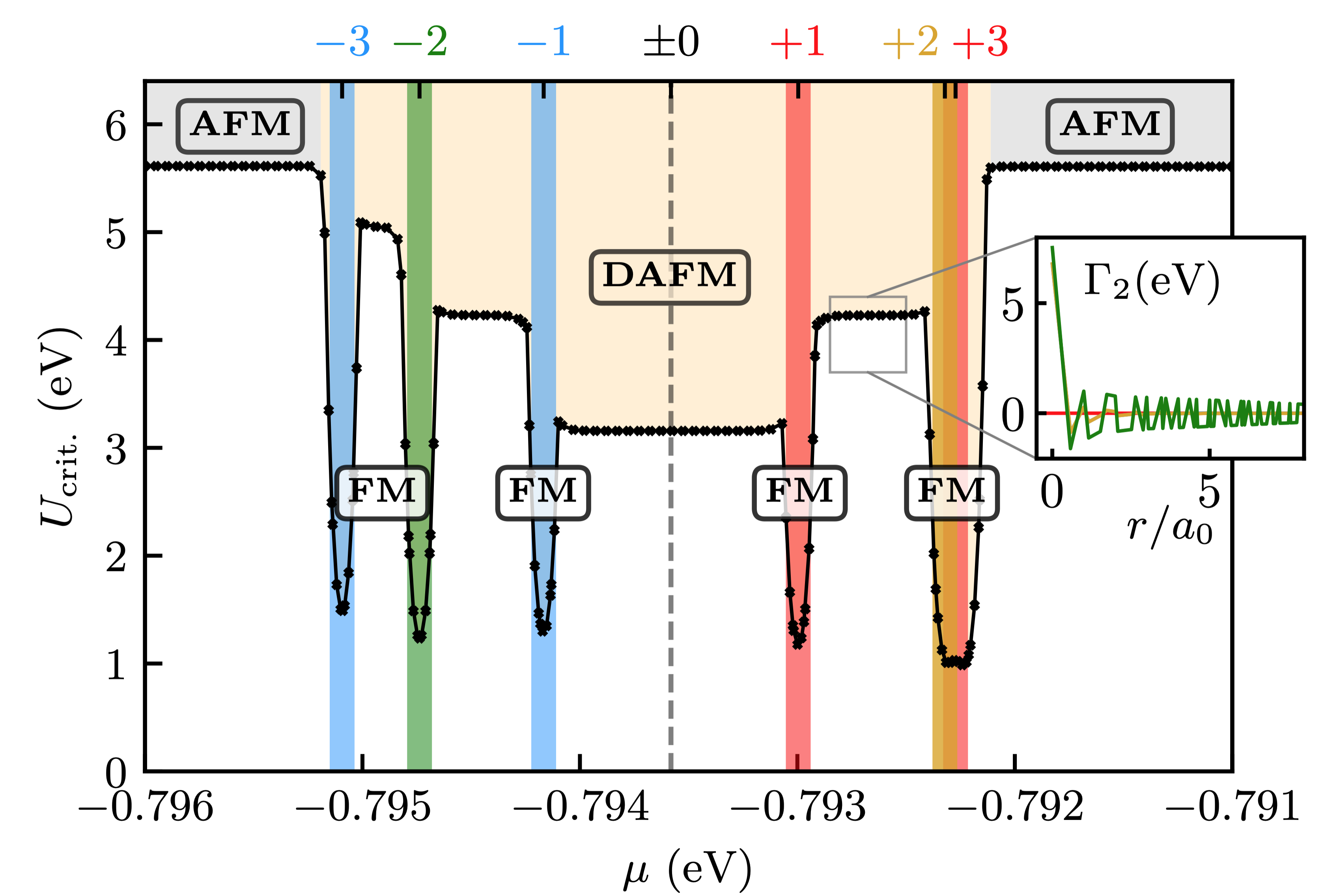}
    \centering
    \caption{Magnetic RPA phase diagram showing the critical onsite interaction strength $U_{\text{crit.}}$ vs. chemical potential $\mu$ in the four flat bands of TBG at $T = 0.03 \, \text{meV}$. The vertical lines indicate the integer fillings $\pm 3, \pm 2, \pm 1$ that show an increased magnetic ordering tendency towards a ferromagnetic state, while away from integer fillings weaker antiferromagnetic tendencies dominate. The boxed abbreviations indicate the type of magnetic ordering: FM - ferromagnetic order, AFM - antiferromagnetic order, DAFM - antiferromagnetic order with real-space node \cite{klebl2019inherited}. The spin-fluctuation interaction (inset) is attractive on nearest-neighbor bonds close to a DAFM instability, while it is purely repulsive near a FM instability.}
    \label{mu_phase}
\end{figure}

\emph{Magnetic instabilities. ---} It is instructive to first analyze the magnetic ordering tendencies suggested by the RPA analysis. The instabilities can be classified according to a generalized Stoner criterion: the effective interaction Eq.~\eqref{eq_veff} diverges, when an eigenvalue of $\hat{\chi}_0$ reaches  $-1/U$ (for details see SM \cite{Note1}). The corresponding eigenvector yields information about the spatial structure of orbital magnetization in the moir\'{e} unit cell \cite{Hertz}. The key result is the phase diagram in Fig. \ref{mu_phase}. It shows the critical interaction strength $U_{\text{crit.}}$ as function of the chemical potential $\mu$ tuned in the four flat bands of TBG. The temperature $T = 0.03 \, \text{meV}$ is fixed for all simulations. For the sake of brevity, we here adapt the classification taken in Ref.~\onlinecite{klebl2019inherited} and label the different leading eigenvectors of $\hat{\chi_0}(0,0)$ according to their real-space profile: FM -ferromagnetic order, AFM - antiferromagnetic order, DAFM - (domain wall) antiferromagnetic order with real-space node (shown in the SM \cite{Note1}). Interestingly, the system shows an increased magnetic ordering tendency towards a ferromagnetic ordered state when the chemical potential is fine-tuned to one of the integer fillings $\pm 3,\pm 2, \pm1$. For partially filled bands, weaker antiferromagnetic patterns prevail. Our results seem to reinforce recent conductance measurements \cite{lu2019superconductors, cao2020tunable} predicting Mott insulator like behavior at charge neutrality and particle-hole asymmetric insulating states at integer fillings. We stress that our eigenvector analysis is not suitable to make any direct quantitative predictions about bands gaps in the electronic spectrum. Nevertheless, the separation of the flat bands from the rest of the spectrum as well as the robustness with the low $U_{\text{crit.}}$ at integer fillings suggest that for these dopings, the ordering should lead to a complete splitting of the flat bands and thus insulating states.


\emph{Unconventional Superconductivity. ---} Next, we proceed with analyzing the spin-fluctuation-induced pairing interaction Eq.~\eqref{eq_veff} in the static, long-wavelength limit.  
At integer fillings $\pm 3,\pm 2, \pm1$, the FM magnetic instability clearly dominates with rather low critical interaction strengths and pairing may not be relevant. 
In between the integer fillings, the DAFM instability dominates the magnetic channel but, depending on the interaction value, may not be strong enough to actually occur. 
Then, below the threshold for the DAFM instability, the attractive pairing channels contained in the effective spin-mediated pairing vertex $\hat{\Gamma}_2$ may induce pairing instabilities at sufficiently low $T$, consistent with experimental measurements, where superconducting regions appear between correlated insulator states located at integer fillings \cite{cao2018a, lu2019superconductors}. 
The real-space dependence of $\hat{\Gamma}_2$ is shown in the inset of Fig.~\ref{mu_phase}.
It is staggered throughout the moir\'e unit cell,
including strong on-site repulsion and nearest-neighbor attraction. Such an alternating interaction is known to allow for pairing interactions for unconventional singlet Cooper pairs living on the bonds bridging the sign changes \cite{Scalapino1995}. On the honeycomb lattice for these band fillings, it is expected to drive spin-singlet $d$-wave pairing states on nearest-neighbor bonds \cite{black2014chiral, Honerkamp2008}, which would now occur in the two graphene sheets of TBG. 
The amplitude of the effective interaction decays exponentially on the c-c bond scale and is largest in the AA regions with only minor contributions in the AB (BA) and DW regions. The same applies for the interlayer interaction, although, importantly, the latter is an order of magnitude smaller than comparable intralayer terms. This indicates that the main pairing will create in-plane Cooper pairs. As the interaction term diverges in the limit $U \to U_{\text{crit}}$, we can effectively control the overall amplitude of $\hat{\Gamma}_2$ by tuning the only free parameter $U$ of our model.

\begin{figure*}
    \centering
    \includegraphics[width = 0.99\textwidth]{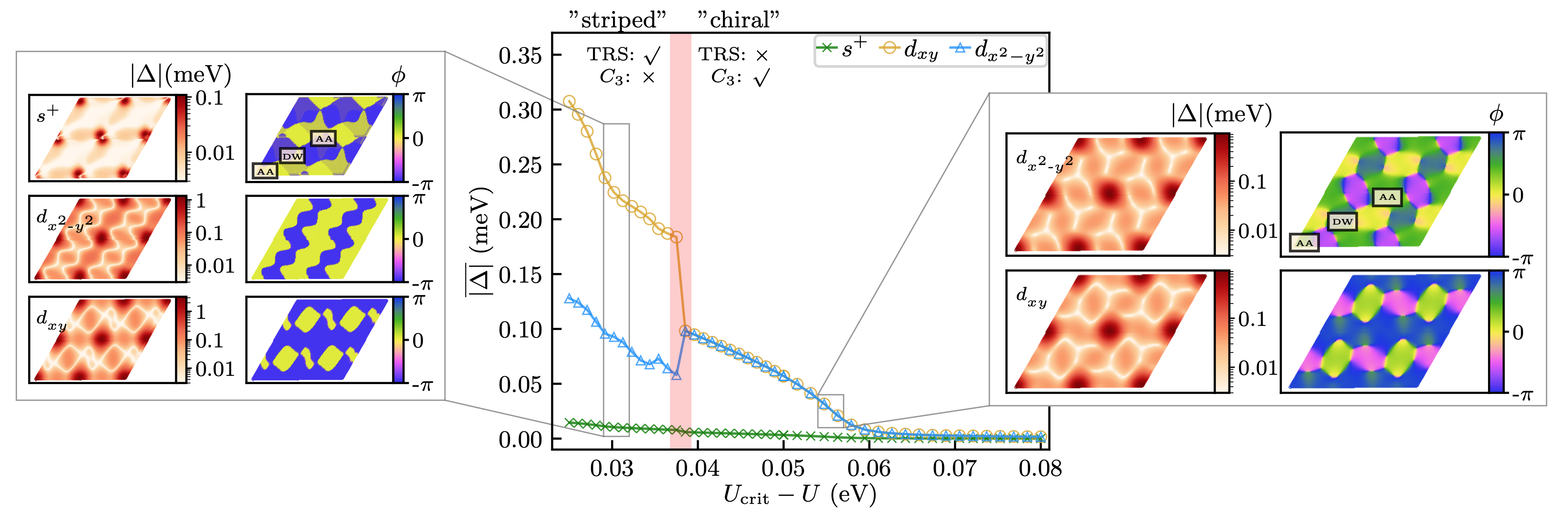}
    \caption{Averaged amplitude of the superconducting order parameter $\overline{\left| \Delta\right |}$ projected onto the form factor basis of nearest-neighbor bonds $s^{+}$, $d_{x^2-y^2}$ and $d_{xy}$ of the original C-atom graphene lattice, as function of the repulsive Hubbard-$U$ for $\mu = -0.7945 \, \text{eV}$ and  $T = 0.03 \, \text{meV}$. The left panel shows the amplitude and phase distribution of the "striped" superconducting phase in the lower layer of TBG, while the right panel depicts the "chiral" phase. The amplitude is strongly enhanced in the AA regions forming superconducting islands, but vanishes in the AB (BA) regions.}
    \label{delta_mean}
\end{figure*}

Based on this argumentation, as a direct consequence of the proximity to a moiré-modulated antiferromagnetic state, we now restrict our mean-field decoupling to spin-singlet configurations
\begin{equation}
\begin{split}
\Delta_{nm}(\bvec{k}) = - \frac{1}{2N} \sum_{\bvec{k}^{\prime} \sigma} &\, \Gamma_{2,nm}(\bvec{q} = \bvec{k} - \bvec{k}^{\prime} , 0) \\
& \times \sigma \langle c_{n \sigma}(\bvec{k}^{\prime}) c_{m \bar{\sigma}}(-\bvec{k}^{\prime}) \rangle_{\text{\tiny MF}} , 
\end{split}
\label{singlet_gap}
\end{equation}
where $\sigma =\pm 1$ is the spin index and $\bar \sigma = - \sigma$.
Based on the fast decay of the pairing interaction on the moiré scale, we can assume that the momentum dependence of the gap function in Eq. \eqref{singlet_gap} is rather weak when viewed in the small moir\'{e} Brillouin zone.
The mentioned $d$-wave pairing will then be encoded in the $nm$-dependence.
Dropping the $\bvec k$-dependence, we solve the resulting Bogoliubov de-Gennes (BdG) equations self-consistently in Nambu space \cite{zhu2016bogoliubov}. {To ensure proper convergence of this large-scale self-consistent calculation, we track the free energy $F \approx  - \sum_n E_n - \sum_{nm}|\Delta_{nm}|^2/\Gamma_{2,nm}$ during each self-consistent cycle and compare the converged  results for different initial gaps (see SM).}
In order to characterize the superconducting order parameter with respect to different pairing channels, we project $\Delta_{nm}$ onto the  real space form-factor basis of the original lattice of carbon atoms,
\begin{equation}
\Delta_{\eta}(n) = \sum_{l} f_{\eta}(\boldsymbol{\delta}_l)(c_{n\uparrow} c_{n + \boldsymbol{\delta}_l \downarrow} - c_{n\downarrow} c_{n + \boldsymbol{\delta}_l \uparrow}),
\label{form_factor}
\end{equation}
where $\eta$ denotes different pairing channels: $s^{+}, d_{xy}$ and $d_{x^2-y^2}$. 
The form factors $f_{\eta}(\boldsymbol{\delta}_l)$ are specific to the pairing channel and result from symmetrizing the bond functions $\boldsymbol{\delta}_l$ with the irreducible representations of the point group $D_{6h}$ of graphene \cite{xu2016competing}. 
Here, we only take the $l=1,2,3$ nearest neighbor C-C bonds $\boldsymbol{\delta}_l$ into account, as they are dominant in terms of attractive interaction strength. 

First, we study the dependence of the order parameter $\Delta_{nm}$ projected on $s$- and $d$-wave as function of the distance to the critical on-site interaction strength $U_{\text{crit.}}$ for fixed chemical potential $\mu = -0.7945 \, \text{eV}$, i.e. between fillings $-2$ and $-1$ and representative for the fillings where stronger $U$ would cause a DAFM instability. When varying $U$, the system makes a first-order phase transition from a "chiral” superconducting phase to a “striped” phase at $U_{\text{crit.}}-U \approx 0.04 \, \text{eV}$ as depicted in Figure \ref{delta_mean}. In the “chiral” phase, the system assumes pure $d+id$ order on the carbon-carbon bond scale, spontaneously breaking time-reversal symmetry (TRS). The order parameter is strongly enhanced in the AA regions, forming superconducting islands that maintain the original $C_3$ symmetry of the normal-state Hamiltonian on the moir\'{e} scale. The amplitude is suppressed and vanishes completely in the AB (BA) regions. At the same time, the phase of the superconducting gap exhibits windings of $2 \pi$ around these regions, giving rise to a vortex-antivortex structure with distinctive signatures in the bond current \begin{equation}
\boldsymbol{J}_{nm} = \frac{e}{i \hbar} \langle c_{n}^{\dagger}t_{nm}c_m - c_m^{\dagger}t_{mn}c_n  \rangle \hat{\boldsymbol{e}}_{nm}.
\label{eq_supercurrent}
\end{equation}
The "striped" phase, however, is characterized by a real $\Delta_{nm}$ and thus restores TRS, but breaks the original $C_3$ symmetry on the moir\'{e} scale. The $s^{+}$ component jumps from zero to a finite value and therefore the phase is of type $s+d$ which was also mentioned in Ref.~\onlinecite{PhysRevB.98.195101}. No vortex structures occur in this phase.
{As the striped phase breaks $C_3$ symmetry but not translational symmetry, it represents a nematic superconducting phase that has been found experimentally in Ref.~\onlinecite{cao2020nematicity}.}
In both phases, we observe a phase shift of $\pi$ between the two graphene sheets as result of the original interlayer repulsion.

\begin{figure}
    \centering
    \includegraphics[width = 0.5\textwidth]{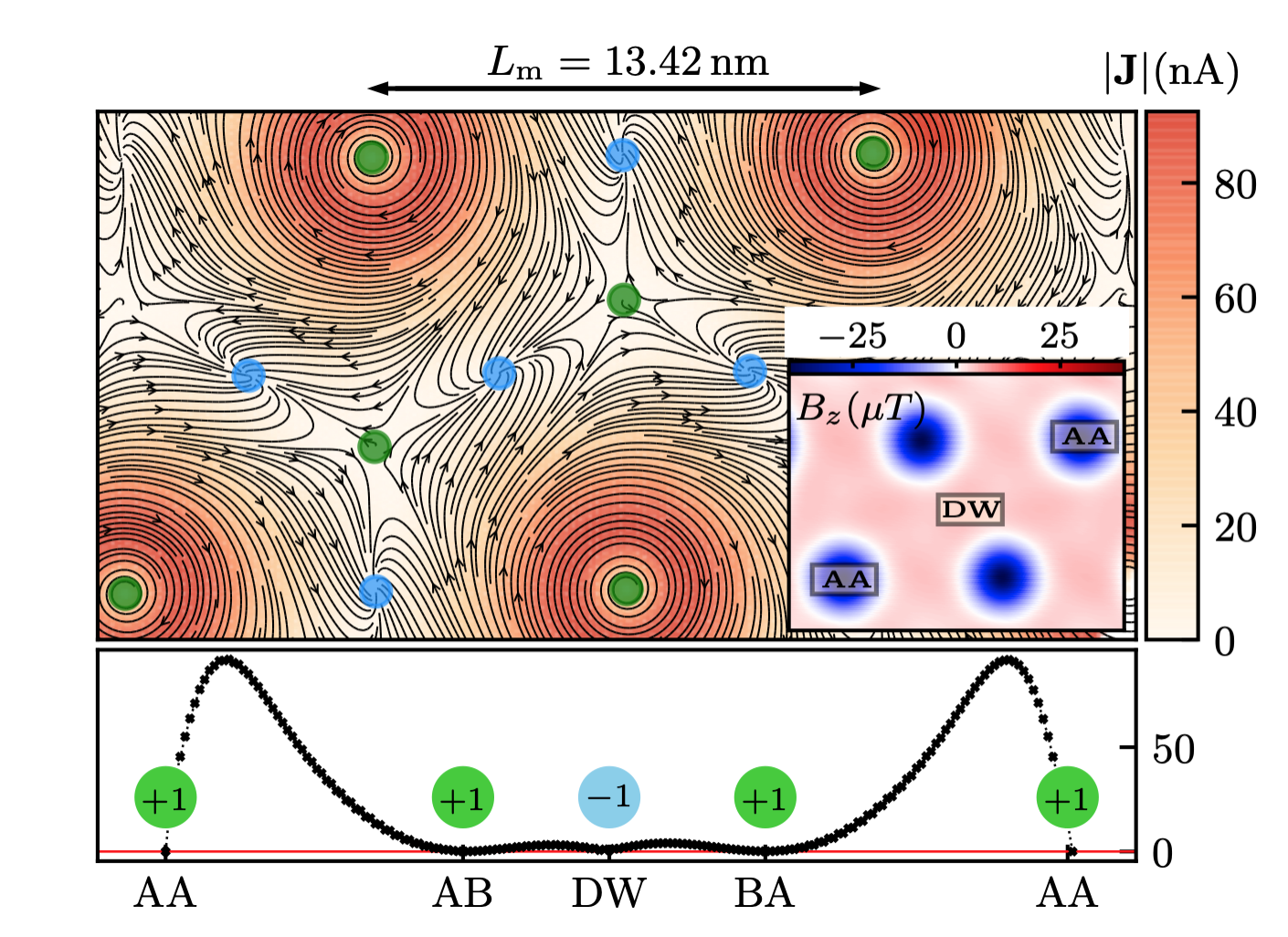}
    \caption{Magnitude and direction of the bond current $\bvec{J}_{nm}$ in the lower layer of TBG for $\mu = -0.7945 \, \text{eV}$,  $T = 0.03 \, \text{meV}$ in the "chiral" superconducting phase of the system. The current pattern indicates vortices in the AA, AB (BA) and DW regions with the vorticity being labelled in the linecut through the moir\'{e} unit cell in the lower panel. The ring current is strongest in the AA regions and the vorticity averages to zero over the moir\'{e} unit cell. The inset shows the magnetic field induced by the supercurrent at a distance corresponding to the interlayer spacing of TBG.}
    \label{current}
\end{figure}

To further analyze the vortex structure appearing in the "chiral" superconducting phase of the system, we compute the layer-resolved quasi-particle bond current \cite{zhu2016bogoliubov} in TBG. 
The current vector field $\bvec J$ shown in Fig.~\ref{current} results from averaging the bond-restricted term $\bvec{J}_{nm}$ over nearest-neighbor sites. The pattern indicates a ring current around the AA regions with vanishing amplitude in the center. The same holds for the AB (BA) and DW region. The current co-rotates in the two graphene sheets, which thus have the same vorticity. In general, the direction of rotation assumed by the system depends on the initial guess for $\Delta_{nm}$ as expected in the context of spontaneous symmetry breaking. 
When starting with time-reversed inital gaps
(suggesting an inverted current) in the two layers, the system converges into a state with higher free energy such that co-rotation is energetically favoured and the magnetic fields created by the supercurrents in both layers add constructively.

In total, we identify six (anti-)vortices per moir\'{e} unit cell: one vortex in the AA region forming a triangular lattice, two vortices in the AB region (honeycomb lattice) and three anti-vortices in the DW regions (kagome lattice), see linecut of Fig. \ref{current} for the direction of rotation. Thus, the total vorticity is zero in the moir\'{e} unit cell. However, the vortices on the triangular lattice of the AA sites dominate the current signal entirely due to the larger order parameter in these regions and the other vortices might not be easily found experimentally.

The spontaneously flowing currents induce a magnetic field that can be calculated by applying Biot-Savart's law. According to our calculations, the magnetic field is of the order of $\mu\mathrm T$ at a distance corresponding to the spacing between the layers. Therefore, the spontaneously formed, current induced magnetic field is within experimental reach of state of the art techniques  \cite{Mitchell_RMP}. Due to aforementioned dominant current amplitude around the AA regions, the more subtle (anti-)vortex structure beyond the  AA region will be very difficult to resolve  in magnetic measurements.

\emph{Conclusion. ---}
We investigated a theoretical scenario in which the superconducting regions found in twisted bilayer graphene are due to electronically mediated paring, in a microscopic model that resolves the carbon-carbon bond length and captures the full $\pi$-bandwidth of the layers. To this end, we used an \textit{ab-initio}-based bandstructure of magic-angle TBG, keeping the many thousand atoms in the unit cell within our modelling. We derived a spin-fluctuation mediated interaction vertex from these microscopic grounds. The phase diagram suggests correlated magnetic and insulating states at integer fillings and attractive electron-mediated interactions between these fillings, giving rise to  unconventional pairing states consistent with recent measurements. A successive mean-field decoupling revealed an inhomogeneous, chiral order parameter on the carbon-carbon scale that is strongly enhanced in the AA region. It exhibits spontaneous supercurrents and magnetic fields in certain parameter regimes that should be measurable signatures in experiment. This shows an exciting extension of the complexity of unconventional superconducting states in these systems: besides breaking the point group symmetry and time reversal by a gap function that winds in wavevector space, the gap function in our case varies strongly and also winds in real space within the moiré unit cell, due to the chirality of the underlying bond pairing.

As next steps, we suggest the experimental scrutiny of the phase-winding signatures of the electronically mediated pairing. These should be addressable by the currents and consequently magnetic fields induced. In theory, one should include phonon-mediated interaction in the "\AA ngstr\"om"-model, e.g. obtained from \textit{ab initio} calculations. This will reveal the interplay and importance of different potential origins of superconductivity \cite{stepanov2019interplay}, which is a highly anticipated avenue of future research. 

\paragraph{Acknowledgment}
The Deutsche Forschungsgemeinschaft (DFG, German Research Foundation) is acknowledged for support through RTG 1995 and under Germany's Excellence Strategy - Cluster of Excellence Matter and Light for Quantum Computing (ML4Q) EXC 2004/1 - 390534769. We acknowledge support from the Max Planck-New York City Center for Non-Equilibrium Quantum Phenomena. Simulations were performed with computing resources granted by RWTH Aachen University under project rwth0554.

\bibliography{moire_sc,DK_ref}

\end{document}


\title{Supplemental Material for \\
Spin-fluctuation-induced pairing in twisted bilayer graphene}
\author{Ammon Fischer} 
\affiliation{Institute for Theoretical Solid State Physics,
RWTH Aachen University, and JARA Fundamentals
of Future Information Technology, 52062 Aachen, Germany}
\author{Lennart Klebl} 
\affiliation{Institute for Theory of Statistical Physics,
RWTH Aachen University, and JARA Fundamentals
of Future Information Technology, 52062 Aachen, Germany}
\author{Carsten Honerkamp}
\affiliation{Institute for Theoretical Solid State Physics,
RWTH Aachen University, and JARA Fundamentals
of Future Information Technology, 52062 Aachen, Germany}
\author{Dante M. Kennes}
\affiliation{Institute for Theory of Statistical Physics,
RWTH Aachen University, and JARA Fundamentals
of Future Information Technology, 52062 Aachen, Germany}
\affiliation{Max Planck Institute for the Structure and Dynamics of Matter, Center for Free Electron Laser Science, 22761 Hamburg, Germany}
\renewcommand\thesection{S~\Roman{section}}

\maketitle

\section{Atomic structure of Twisted Bilayer Graphene}

The atomic structure of twisted bilayer graphene (TBG) consists of two super-imposed graphene layers rotated by an angle $\theta$ and separated by a distance of $d_0 = 0.335$ nm. The axis of rotation is chosen to intersect two vertically aligned carbon atoms when starting from purely AA-stacked bilayer, i.e. two perfectly overlapping honeycomb lattices. Labelling the upper (lower) monolayer of TBG by $u$($l$), their two-dimensional lattice vectors are 
\begin{align}
\begin{split}
\bvec{a}_1^l &= a_0(\text{cos}(\pi/6) , -\text{sin}(\pi/6) ) \\
\bvec{a}_2^l &= a_0(\text{cos}(\pi/6) , \ \ \text{sin}(\pi/6) ) \\
\bvec{a}_i^u &= R(\theta)\bvec{a}_i^l,  
\end{split}
\label{lattice_vec}
\end{align}
where $R(\theta)$ represents a rotation by an angle $\theta$ and $a_0 = 0.246 \,  \text{nm}$ is the lattice constant of graphene, which should not be confused with the carbon-carbon bond length of $a_{\text{cc}} = a_0/\sqrt{3}$. The so defined primitive cell is di-atomic and contains two inequivalent sites with basis vectors
\begin{align}
\begin{split}
\bvec{b}_1^{u(l)} &= (0,0) \qquad \qquad \qquad \quad \text{(A site)}, \\
\bvec{b}_2^{u(l)} &= \frac{1}{3} \bvec{a}_1^{u(l)}  + \frac{1}{3} \bvec{a}_2^{u(l)} \qquad  \, \, \,  \text{(B site)}.
\end{split}
\label{lattice_basis}
\end{align}

\begin{figure*}
    \includegraphics[width=\textwidth]{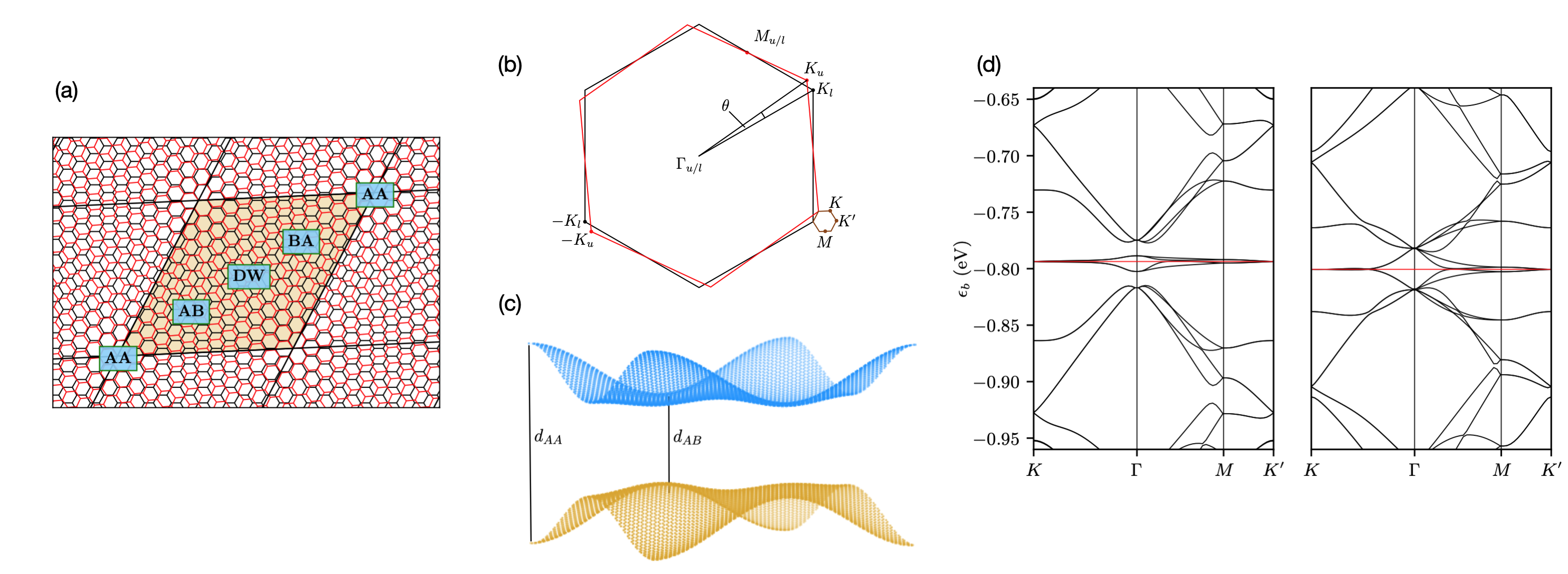}
    \centering
    \caption{\textbf{Atomic structure of twisted bilayer graphene.} \textbf{(a)} Atomic structure of twisted bilayer graphene with twist angle $\theta = 5.09^{\circ}$ corresponding to $(m,n) = (6,7)$. The blue labels indicate characteristic stacking patterns emerging throughout the moir\'{e} unit cell. \textbf{(b)} Downfolding of the (mini-) Brillouin zone of TBG (brown hexagon). \textbf{(c)} Corrugation effects in TBG. \textbf{(d)} Band structure of magic-angle TBG with twist angle $\theta = 1.05^{\circ}$ corresponding to $(m,n) = (31,32)$. The low-energy window around charge neutrality (red line) is modified significantly when corrugation of the two graphene sheets is taken into account (left panel). This is reflected in the formation of four flat bands (two-fold spin degeneracy) around charge neutrality that have a bandwidth of $\approx 15 \, \text{meV}$ and are separated from the rest of the spectrum.}
    \label{fig:atomic_structure}
\end{figure*}

The choice of basis vectors and the type of rotation around an AA site restricts the symmetry of TBG to point group $D_3$. The latter contains a threefold in-plane rotation $C_3 = C_{3z}$ around the z-axis and a twofold out-of-plane rotation $C_2 = C_{2y}$. \\
The lattice geometry of TBG, as defined so far, is not periodic in general \cite{PhysRevX.8.031087, moon2013optical, dos2012continuum} since the periods of the two graphene layers are incommensurate for arbitrary twist angles. A finite unit cell can only be constructed for some discrete angles satisfying the condition
\begin{align}
\text{cos}(\theta) =  \frac{1}{2} \frac{m^2+n^2+mn}{m^2+n^2+mn}
\label{theta_moire}
\end{align}
with $(m,n)$ being positive integers. In this case, the twisted bilayer graphene forms a moir\'{e} pattern, see Fig.~\ref{fig:atomic_structure} (a), containing $N(m,n) = 4(m^2+n^2+mn)$ atoms with superlattice vectors
\begin{align}
\begin{split}
\bvec{L}_1 &= m \bvec{a}_1^l + n \bvec{a}_2^l  = n \bvec{a}_1^u + m \bvec{a}_2^u  \\
\bvec{L}_2 &= R(\pi/3)\bvec{L}_1.  
\end{split}
\label{moire_vec}
\end{align}

Magic-angle twisted bilayer graphene has a twist angle of $\theta = 1.05^{\circ}$ corresponding to the the integers $(m,n) = (31,32)$. The so defined structure contains $N = 11908$ carbon atoms in the moir\'{e} unit cell.
When discussing the geometric structure of TBG, atomic relaxation effects play an important role and may modify the low-energy physics of the system significantly. From experiments using transmission electron microscopy (TEM)  \cite{PhysRevB.48.17427} as well as from structural optimization studies using density functional theory \cite{uchida2014} it is known that the interlayer distance between the two layers varies over the moir\'{e} unit cell. The interlayer spacing takes its maximum value $d_{AA} = 0.360 \, \text{nm}$ in the AA regions and its minimal value $d_{AB} = 0.335 \, \text{nm}$ in the AB regions. Intermediate spacings $d(\bvec{r})$ are obtained by using an interpolation suggested by \cite{PhysRevX.8.031087, uchida2014}
\begin{equation}
d(\bvec{r}) = d_0  + 2d_1\sum_{i = 1}^3 \text{cos} \left ( \bvec{G}_i  \cdot \bvec{r} \right), 
\label{corr}
\end{equation}
where the vector $\bvec{r}$ points to a carbon atom in the moir\'{e} unit cell and $\bvec{G}$ are the reciprocal lattice vectors obtained from Eq. \eqref{moire_vec}. Furthermore, the constants $d_0 = \frac{1}{3}(d_{AA} + 2d_{AB})$ and $d_1= \frac{1}{9}(d_{AA} - d_{AB})$ are defined such to match the distances in the AA and AB regions. In order to preserve the $D_3$ symmetry of the system, the corrugation must be applied symmetric to both layers as depicted in Fig.~\ref{fig:atomic_structure} (c).

\section{Atomistic Tight-binding Hamiltonian}

The eigenenergies and eigenfunctions of non-interacting TBG are obtained using a single-orbital tight-binding Hamiltonian for the $p_z$-orbitals of the carbon atoms \cite{PhysRevX.8.031087, moon2013optical, sboychakov2015electronic}:
\begin{equation}
H = \sum_{\bvec{R}, \bvec{R'}} \sum_{i,j, \sigma} t(\bvec{R} + \bvec{r}_i - \bvec{R'} - \bvec{r}_j) c_{\bvec{R}, \bvec{r}_i, \sigma}^{\dagger} c_{\bvec{R'}, \bvec{r}_j, \sigma}.
\label{tb_ham}
\end{equation}
For our microscopic approach we account for the full $\pi$-band spectrum of magic-angle TBG, keeping all $N=11908$ bands under consideration.
In the following, we label the supercell vector with $\bvec{R}$ and the position vector of site $i$ in the corresponding moir\'{e} unit cell with $\bvec{r}_i$. Hence, the operator $c_{\bvec{R}, \bvec{r}_i, \sigma}^{\dagger}$ creates an electron with spin $\sigma = {\uparrow, \downarrow}$ in the $p_z$-orbital of site $i$, whereas $c_{\bvec{R}, \bvec{r}_i, \sigma}$ destroys an electron with the same quantum numbers. The transfer integral between orbitals at site $i$ and $j$, separated by the vector $\bvec{d}$ can be written in Slater-Koster form \cite{PhysRevX.8.031087}
\begin{align}
\begin{split}
t(\bvec{d}) &= t_{\parallel}(\bvec{d}) + t_{\bot}(\bvec{d}) \\
t_{\parallel}(\bvec{d}) &= -V_{pp \pi}^0\text{exp} \left(-\frac{d-a_{\text{cc}}}{\delta_0}\right) \left[1 - \left(\frac{d^z}{d} \right)^2 \right] \\
t_{\bot}(\bvec{d}) &= -V_{pp \sigma}^0\text{exp} \left(-\frac{d-d_0}{\delta_0}\right)\left[\frac{d^z}{d} \right]^2.
\end{split}
\label{tb_hop}
\end{align}
Here, $d^z = \bvec{d} \cdot \bvec{e}_z$ points perpendicular to the graphene sheets and $d_0 = 1.362 \, a_0$ is the vertical spacing of graphite. The term $V_{pp \sigma} =  0.48 \ \text{eV}$ describes the interlayer hopping between atoms in different monolayers of TBG, while $V_{pp \pi} =   - 2.7 \ \text{eV}$ models the intralayer hopping amplitude between neighboring atoms in a single graphene sheet. The parameters are fitted to data of first principle calculations to match the dispersion of mono- and bilayer graphene \cite{PhysRevX.8.031087}. This ensures in particular that TBG behaves locally similar to graphene with the overall structure being modulated by the moir\'{e} pattern. The parameter $\delta_0 = 0.184 \, a_0$ determines the decay length of the transfer integral and is chosen such that the nearest-neighbor intralayer hopping reduces to $0.1 V_{pp\pi}$. For numerical calculations it is therefore sufficient to truncate hopping terms for $r_{ij} > 4 \, a_0 $ as $t(\bvec{r}_{ij}) < 10^{-4}$ in these regimes.

To construct the full non-interacting Hamiltonian of the periodic system, we define the Bloch wave basis by a Fourier transform to mini-Brillouin zone (MBZ) momentum $\bvec{k}$, which is depicted in Fig.~\ref{fig:atomic_structure} (b),
\begin{align}
\begin{split}
c_{\bvec{k}, \bvec{r}} &= \frac{1}{\sqrt{N}} \sum_{\bvec{R}} e^{i \bvec{k} \bvec{R}} c_{\bvec{R}, \bvec{r}} \\
c_{\bvec{R}, \bvec{r}} &= \frac{1}{\sqrt{N}} \sum_{\bvec{k}} e^{-i \bvec{k} \bvec{R}} c_{\bvec{k}, \bvec{r}}.
\end{split}
\label{bloch_op}
\end{align}

The spin index is omitted for simplicity. Note that the moir\'{e} Fourier transform as defined above couples momenta $\bvec{k}$ and superlattice vectors $\bvec{R}$. The latter describe the spatial extent of the moir\'{e} unit cell, which in the case of magic-angle TBG is given by $L = |\bvec{L}_{1,2}| = 13.42 \, \text{nm}$.

Inserting these expressions into Eq.\eqref{tb_ham} renders the unperturbed Hamiltonian block-diagonal in momentum space
\begin{align}
\begin{split}
H_0 &= \sum_{\bvec{k}}\sum_{\bvec{R} - \bvec{R'}} \sum_{i,j, \sigma} t(\bvec{R} + \bvec{r}_i - \bvec{R'} - \bvec{r}_j) \\ & \qquad \qquad \qquad \quad \times e^{-i\bvec{k} \cdot(\bvec{R}-\bvec{R}^{\prime})} c_{\bvec{k}, \bvec{r}_i, \sigma}^{\dagger} c_{\bvec{k}, \bvec{r}_j, \sigma}^{\phantom \dagger} \\
&= \sum_{\bvec{k}} \sum_{i,j, \sigma} \left  [ H_0(\bvec{k}) \right ]_{ \bvec{r}_i, \bvec{r}_j} c_{\bvec{k}, \bvec{r}_i, \sigma}^{\dagger} c_{\bvec{k}, \bvec{r}_j, \sigma}^{\phantom \dagger}. 
\end{split}
\label{BH}
\end{align}
The matrix $ \left  [ H_0(\bvec{k}) \right ]_{ \bvec{r}_i, \bvec{r}_j}$ can be diagonalized in orbital space $(i,j)$ for each $\bvec{k}$ to obtain the bandstructure $\epsilon_b(\bvec{k})$ and orbital-to-band transformation $u_{\bvec{r}}^{b}(\bvec{k})$, $b = 1 .. N$: 
\begin{align}
H_0 &= \sum_{\bvec{k}, b} \epsilon_b(\bvec{k}) \gamma_{\bvec{k}, b}^{\dagger} \gamma_{\bvec{k}, b}^{\phantom \dagger} \qquad \text{with} \ \   \gamma_{\bvec{k}, b} = u_{\bvec{r}}^{b}(\bvec{k}) c_{\bvec{k}, \bvec{r}}.
\label{o2b}
\end{align}
Since magic-angle TBG contains $N=11908$ atoms in the moir\'{e} unit cell, care must be taken when treating the system numerically.

\section{Magnetic instabilities and Stoner criterion}
\subsection{Stoner Criterion}
In the manuscript, we study magnetic instabilities in TBG described by short-ranged Coulomb interactions. To this end, we follow Ref.~\cite{klebl2019inherited} and employ a repulsive Hubbard term for electrons with opposite spin $\sigma$ with $\overline{\sigma} = -\sigma$ residing on the same carbon site
\begin{equation}
H_{\text{int}} = \frac{1}{2} U \sum_{\bvec{R},i, \sigma} n_{\bvec{R}, \bvec{r}_i, \sigma} n_{\bvec{R}, \bvec{r}_i, \overline{\sigma}}.
\label{hubbard}
\end{equation}
To treat the interacting system in a pertubative manner, we define the free Matsubara Green's function in orbital-momentum space as
\begin{equation}
g_{\bvec{r}, \bvec{r}^{\prime}} (i\omega, \bvec{k}) = \sum_{b} u_{\bvec{r}}^{b}(\bvec{k}) (i \omega -e_b(\bvec{k}))^{-1} u_{\mathbf{r^{\prime}}}^{b*}(\bvec{k}). 
\label{green}
\end{equation}
We then calculate the renormalized interaction in the spin channel within the random-phase approximation (RPA) to analyze the electronic instabilities mediated by spin-fluctuation exchange between electrons to high order in the bare coupling $U$. Since the initial short-ranged interaction vertex $U$ has no momentum and frequency dependence, the full susceptibility of the system can be approximated by $\hat{\chi}^{\text{RPA}}(q) = \hat{\chi}_0(q)/[1+U \hat{\chi_0}(q)]$. Here, we use the multi-index qunatum number $q = (\bvec{q}, i \omega)$ and indicate matrices of dimension $N \times N$ with an hat symbol, e.g. $\hat{\chi} = \chi_{\bvec{r}, \bvec{r}^{\prime}}$. Magnetic instabilities may subsequently be classified according to a generalized Stoner criterion: The effective (RPA) interaction diverges, when the smallest eigenvalue $\lambda_0$ of $\hat{\chi}(q)$ reaches $-1/U$, marking the onset of magnetic order for all interaction strengths  $U \geq U_{\text{crit.}} = -1/\lambda_0$. The corresponding eigenvector $v^{(0)}(q)$ is expected to dominate the spatial structure of orbital magnetization.

In this letter, we study magnetic instabilities with emphasis on the static, long-wavelength limit $(\bvec{q}, i \omega \to 0)$ on the moir\'{e} scale. The latter limit proves to contain the relevant physics when starting with local repulsive interaction. We stress again that momenta $\bvec{q}$ are related via moir\'{e} Fourier transform Eq. \eqref{bloch_op} to the superlattice vectors $\bvec{R}$. The RPA susceptibility predicts spin correlations at length scales intermediate to the c-c bond scale and moir\'e length scale, thus being described by orderings at $\bvec q=0$. The system shows the same order in all moir\'{e} unit cells with variable correlations present on the c-c bond scale.


\subsection{Spin susceptibility}
For analyzing the magnetic properties of the system on the RPA level, it is therefore sufficient to compute the free polarization function $\hat{\chi}_0(q)$ defined as
\begin{equation}
\chi_{0_{\bvec{r}, \bvec{r}^{\prime}}}(\bvec{q}, i\omega) = \frac{1}{N \beta} \sum_{\bvec{k}, \omega^{\prime}} g_{\bvec{r}, \bvec{r}^{\prime}}(i\omega^{\prime}, \bvec{k})g_{\bvec{r}^{\prime}, \bvec{r}}\left ( i(\omega^{\prime}+ \omega \right ), \bvec{k}+ \bvec{q}).
\label{chi0}
\end{equation}
The Matsubara summation occuring in Eq. \eqref{chi0} can be evaluated analytically yielding the well-known Lindhard function for multi-orbital systems
\begin{equation}
\begin{split}
\label{chi0_mats}
\chi_{0_{\bvec{r}, \bvec{r}^{\prime}}}(\bvec{q}, i\omega) &=   \frac{1}{N} \sum_{\bvec{k}, b, b^{\prime}} \frac{n_F(\epsilon_{b^{\prime }}(\bvec{k})) - n_F(\epsilon_b(\bvec{k}+\bvec{q})) }{i \omega + \epsilon_{b^{\prime}}(\bvec{k}) - \epsilon_b(\bvec{k}+\bvec{q}) } \\ 
&\times u_{\bvec{r}}^{b^{\prime}}(\bvec{k})  u_{\bvec{r}^{\prime}}^{b^{\prime}*}(\bvec{k}) u_{\bvec{r}}^{b*}(\bvec{k}+\bvec{q}) u_{\bvec{r}^{\prime}}^{b}(\bvec{k}+\bvec{q}),
\end{split}
\end{equation}
where $n_F(\epsilon) = (1+e^{\beta \epsilon})^{-1}$ is the Fermi function. While the analytical evaluation of the Matsuabra sum occuring in Eq. \eqref{chi0} is the standard procedure for systems containing only few atoms in the unit cell, this approach is destined to fail in our atomistic approach as it scales like $\mathcal{O}(N^4)$. For magic-angle TBG with $N = 11908$ atoms in the moir\'{e} unit cell, it is more efficient to compute the Matsubara sum in Eq. \eqref{chi0} numerically over a properly chosen frequency grid and in each step compute Hadamard products of band-summed non-local Green's functions $g_{\bvec{r}, \bvec{r}^{\prime}}(i\omega^{\prime}, \bvec{k})g^{T}_{\bvec{r}, \bvec{r}^{\prime}}\left ( i(\omega^{\prime}+ \omega \right ), \bvec{k}+ \bvec{q})$. The expression then scales like $\mathcal{O}(N^3 N_{\omega})$ with $N_{\omega}$ being the number of fermionic frequencies needed to achieve proper convergence. To this end, non-linear mixing schemes \cite{ozaki2007} are proven to outperform any linear summation such that we only need to sum over $N_{\omega} \approx 1000 $ frequencies when accessing temperatures down to $T = 0.03 \, \text{meV}$. The momentum sum occuring in Eq. \eqref{chi0} is evaluated over 24 $\bvec{k}$ points in the MBZ using a momentum meshing proposed by Cunningham et al. \cite{cunningham1974}. In particular, we checked that the results are sufficiently converged when taking an denser mesh into account. 

\begin{figure*}
    \includegraphics[width=\textwidth]{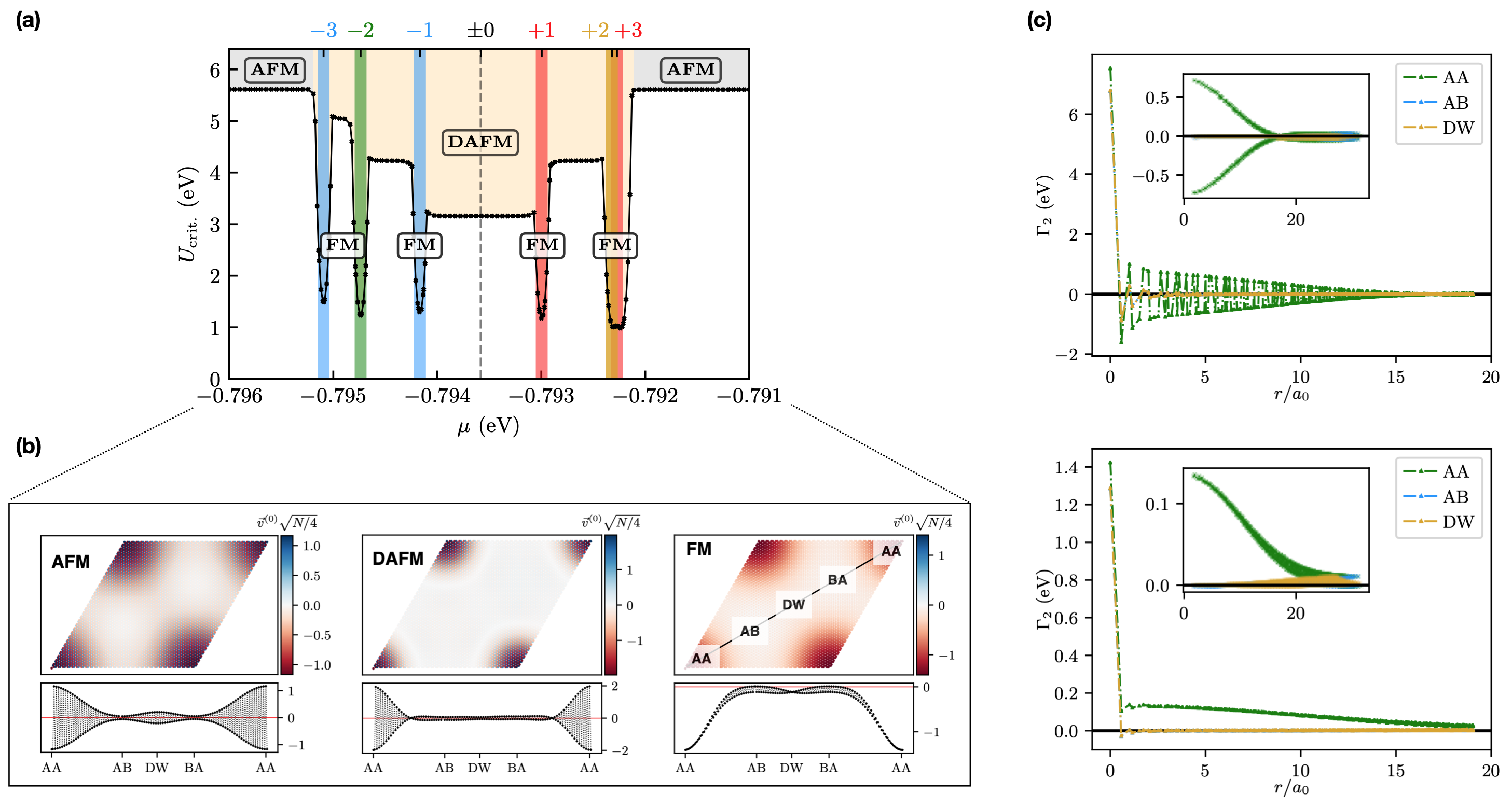}
    \centering
    \caption{\textbf{Spin correlations in magic-angle TBG.}
    \textbf{(a)}
    Magnetic RPA phase diagram showing the critical onsite interaction strength
    $U_{\text{crit.}}$ vs. chemical potential $\mu$ in the four flat bands of TBG at $T = 0.03 \,
    \text{meV}$. The vertical lines indicate the integer fillings $\pm 3, \pm 2, \pm 1$ that show
    an increased magnetic ordering tendency towards a moir\'{e}-modulated ferromagnetic state,
    while away from integer fillings weaker antiferromagnetic tendencies dominate.
    \textbf{(b)}
    Spatial distribution of the leading eigenvector of the RPA analysis on the carbon-carbon bond
    scale in the moir\'{e} unit cell. For simplicity, only the lower layer of TBG is shown as well
    as a linecut through the unit cell. The leading eigenvector of the DAFM instabilities is
    staggered throughout the moir\'{e} unit cell with strongest weight in the AA regions as
    depicted in the linecut in (b).
    \textbf{(c)}
    Effective spin-fluctuation mediated pairing
    vertex $\Gamma_2(q=0)$ close to a DAFM (upper panel) and FM (lower panel) magnetic instability.
    In the former case, the pairing vertex is staggered in real-space with strong on-site
    repulsion and nearest-neighbor attraction. Real-space profiles are shown starting from an atom
    located in the AA, AB or DW region of TBG, respectively. }
    \label{fig:correlation}
\end{figure*}

\subsection{Leading Instabilities}
In the manuscript, we classify the different leading eigenvectors of the RPA analysis according to their real-space profile in the moir\'{e} unit cell following the nomenclature introduced in Ref.~\cite{klebl2019inherited}. The three potential ground states of the interacting system at $T = 0.03 \, \text{meV}$ are depicted in Fig.~\ref{fig:correlation} (b): (i) AFM: moir\'{e}-modulated antiferromagnetic phase on the carbon-carbon bond scale ("\AA ngstr\"om"-scale) with increased weight in the AA regions, (ii) DAFM: moir\'{e}-modulated antiferromagnetic phase with opposite signs between the AA and AB regions that becomes visible as a node in the absolute value of the order parameter, (iii) FM: moir\'{e}-modulated ferromagnetic phase (FM) that exhibits the same overall sign in the moir\'{e} unit cell. The choice of the temperature $T = 0.03 \, \text{meV}$ should provide resolution of the flat energy bands in TBG as extensively discussed in Ref.~\cite{klebl2019inherited}. As long as $T \gg \mathcal{O}(1 \, \text{meV})$ the flat band physics is not resolved and the system inherits magnetic order from purely AA and AB stacked bilayer graphene. For $T \approx \mathcal{O}(1 \, \text{meV}) \approx 10 \, \text{K}$, significant deviations due to the flat bands occur, leading to the plethora of magnetic phases described above.

\section{Self-consistent Bogoliubov de-Gennes equations for TBG}
\subsection{Fluctuation-Exchange approximation}
For interaction values $U < U_{\text{crit}}$ the system is in the paramagnetic regime and the magnetic instabilities prescribed by the RPA analysis are not strong enough to actually occur. In this regime, spin and charge fluctuations contained in the transverse and longitudinal spin channel can give rise to an effective interaction between electrons that may lead to the formation of Cooper pairs. The leading RPA diagrams to the irreducible singlet particle-particle scattering vertex $\hat{\Gamma}_2(\bvec{q},\nu)$ are captured within the fluctuations-exchange approximation (FLEX) \cite{berk1966effect, romer2012local} 
\begin{equation}
\hat{\Gamma}_2(\bvec{q}) = \hat{U}  - \frac{U^2\hat{\chi}_0(\bvec{q})}{1+U\hat{\chi}_0(\bvec{q})}  + \frac{U^3\hat{\chi}_0^2(\bvec{q})}{1-U^2\hat{\chi}_0^2(\bvec{q})}.
\label{eq_veff}
\end{equation}
As in the previous section, we only consider the static long-wavelength limit $(\bvec{q}, i \omega \to 0)$ and thus focus on the pairing structure on the carbon-carbon bonds within the moir\'{e} unit cell. To this end, the real-space profile of the effective interaction $\hat{\Gamma}_2$ for different chemical potentials is shown in Fig.~\ref{fig:correlation} (c). As mentioned in the manuscript, the interaction vertex is staggered through the moiré unit cell close to a DAFM/AFM instability, opening the door for unconventional singlet Cooper pairs in the two graphene sheets of TBG. In particular, the interlayer interaction strength is an order of magnitude smaller than comparable intralayer terms. This indicates that the main pairing will create in-plane Cooper pairs. In the manuscript, we thus only visualize the projection of the superconducting order parameters on the in-plane form factor basis of each graphene sheet, i.e. a layer-resolved representation.

\subsection{Self-consistent BdG Formalism}
\begin{figure}
    \includegraphics[width=0.3\textwidth]{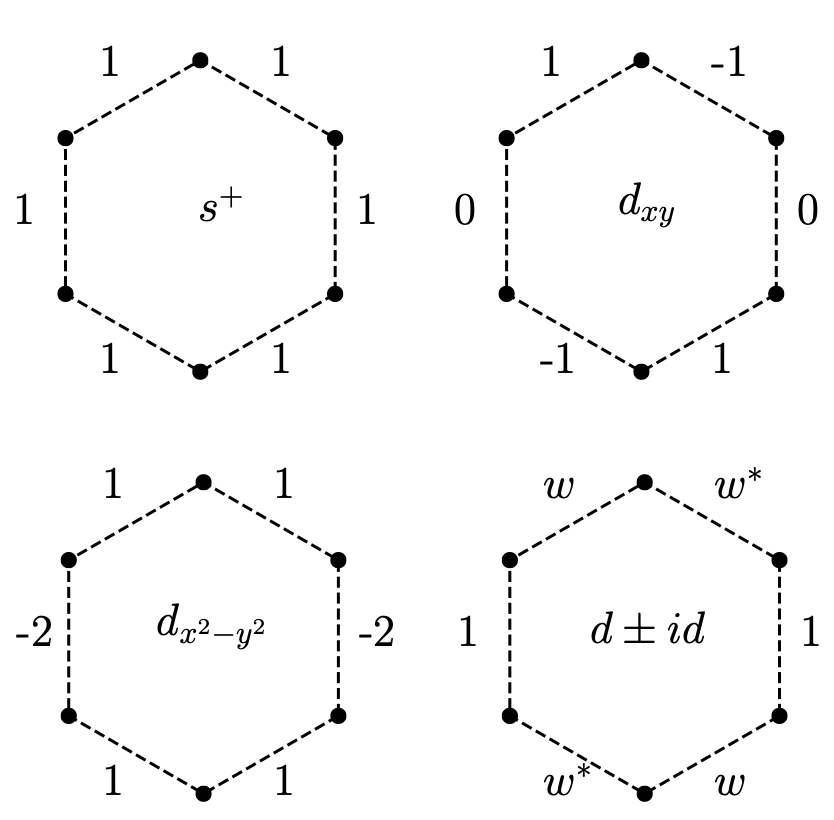}
    \centering
    \caption{Form factors for different nearest-neighbor (singlet) pairing channels on the honeycomb lattice. The complex linear combination $d\pm id$ is characterized by the phase factor $w = \text{exp}(\pm i\, 2 \pi / 3)$. }
    \label{fig:form_factor}
\end{figure}

In the next step, we analyze the effective particle-particle scattering vertex Eq.~\eqref{eq_veff} using a mean-field decoupling to extract pairing symmetries and spatial distribution of the superconducting order parameter. In the static long-wavelength limit $(\bvec{q}, i \omega \to 0)$, we may hence neglect the momentum dependence of the gap parameter and  effectively solve a one-unit cell system with periodic boundary conditions. While this approach does not take correlations between different moir\'{e} unit cells into account, it allows for all pairing contributions from within the moir\'{e} unit cell.
Due to the proximity to the anti-ferromagnetic ordered state, we restrict the mean-field decoupling to spin-singlet configurations that are symmetric under the exchange of spatial indices
\begin{equation}
\begin{split}
\Delta_{ij}= - \frac{1}{2} \left [ \Gamma_2 (q=0) \right]_{ij} \langle c_{i \uparrow} c_{j \downarrow} -  c_{i \downarrow} c_{j \uparrow} \rangle_{\text{\tiny MF}}.
\end{split}
\label{singlet_gap}
\end{equation}
The expectation value $\langle \cdot \rangle_{\text{\tiny MF}}$ can be calculated by diagonalizing the resulting mean-field Hamiltonian $H_{\text{MF}}$ in Nambu-space using a Bogoliubov de-Gennes transformation \cite{zhu2016bogoliubov}
\begin{equation}
\centering
\begin{split}
H_{\text{MF}} &=  \psi^{\dagger}
\begin{pmatrix} \hat{H}_0 & \hat{\Delta}\\ \hat{\Delta}^{\dagger} & -\hat{H}_0  \\\end{pmatrix}
\psi \\
\langle c_{i \uparrow} c_{j \downarrow} -  c_{i \downarrow} c_{j \uparrow} \rangle_{\text{\tiny MF}} &= 
\sum_n \left (u_i^n v_j^{n*} + u_j^n v_i^{n*} \right)\text{tanh} \left(\frac{E^n}{2T}\right ).
\end{split}
\label{bdg}
\end{equation}
Here, $u_i^n \, (v_i^n)$ are the particle (hole) amplitudes of the BdG quasi-particles resulting from the diagonalization of the Hamiltonian in Eq. \eqref{bdg} 
\begin{equation}
\begin{split}
H_{\text{MF}} &= \left ( \hat{U} \psi \right )^{\dagger} \begin{pmatrix} \hat{E} &0 \\ 0 & -\hat{E}  \end{pmatrix} \left ( \hat{U} \psi \right )^{\phantom \dagger} \\
\hat{U} &= \begin{pmatrix} \hat{u} & \hat{v} \\ -\hat{v}^*& \hat{u} \\ \end{pmatrix}, 
\end{split}
\label{bdg_particle_hole}
\end{equation}
and $\psi =  (c_{1 \uparrow}^{\phantom \dagger}, ... , c_{N \uparrow}^{\phantom \dagger},  c_{1 \downarrow}^{\dagger}, ... , c_{N \downarrow}^{\dagger})^{T}$ is the $2N$-component Nambu vector. $\sum_n$ denotes a sum over the positive quasi-particle energies $E_n>0$ and $\hat{E}$ is the corresponding diagonal matrix. To solve this set of self-consistent equations we start with an initial guess for $\Delta_{ij}$ and iterate until convergence is achieved using a linear mixing scheme to avoid any bipartite solutions. Since the atomic arrangement is highly inhomogeneous in the moir\'{e} unit cell, we track the free energy during each self-consistency cycle and for different initial configurations to ensure proper convergence of the algorithm into the actual global minimum. The free energy of the system in the low temperature regime reads
\begin{equation}
F = E-TS \approx E_g - \sum_{n} E_n - \sum_{ij} \frac{|\Delta_{ij}|^2}{\Gamma_{2, ij}}
\label{free_energy}
\end{equation}
where $E_g = 2 \sum_n E_n n_F(E_n)$ is the excitation energy of the quasi-particles. 
The different initial configurations are chosen to transform according to the irreducible representations of the $D_{6h}$ point group of the honeycomb lattice. This procedure aligns with the insights from the previous paragraph that the spin-fluctuation mediated pairing vertex Eq.~\eqref{eq_veff} will create in-plane Cooper pairs with strongest pairing amplitude living on the nearest-neighbor bonds of the two single graphene sheets. The phase factors of the different nearest-neighbor pairing channels are shown in Fig.~\ref{fig:form_factor}.

\subsection{Supercurrent and magnetic field}

To characterize the different superconducting phases of the system, we compute the layer resolved quasi-particle bond current in TBG \cite{zhu2016bogoliubov}
\begin{equation}
\boldsymbol{J}_{nm} = \frac{e}{i \hbar} \langle c_{n}^{\dagger}t_{nm}c_m - c_m^{\dagger}t_{mn}c_n  \rangle \hat{\boldsymbol{e}}_{nm}.
\label{eq_supercurrent}
\end{equation}
In the atomistic approach presented here, the quasi-particle current $\bvec{J}_{nm}$ is only defined between two carbon atoms residing at sites $\bvec{r}_{n(m)}$ in the moir\'{e} unit cell. Therefore, we take an amplitude-weighted average of neighboring bonds to arrive at a vector field representation $\bvec{J}$ as shown in Fig. 3 in the manuscript
\begin{equation}
\bvec{J}(\bvec{r}_n) = \frac{1}{3} \sum_{\langle m \rangle } \bvec{J}_{nm} \hat{\bvec{e}}_{nm}.
\label{current_averge}
\end{equation}
Here, $\hat{\bvec{e}}_{nm}$ points from the atom at position $\bvec{r}_n$ to its three nearest-neighbors $\bvec{r}_m$ on each graphene sheet. In particular, the current amplitude is negligible at distances exceeding nearest-neighbor atoms and thus an average over nearest-neighbors is sufficient.

The spontaneously flowing currents of quasi-particles induce a magnetic field that can be calculated by applying the Biot-Savart law
\begin{equation}
\bvec{B}(\bvec{r}) = \frac{\mu_0}{4 \pi} \int \bvec{J}(\bvec{r}) \times \frac{\bvec{r}-\bvec{r}^{\prime}}{|\bvec{r}-\bvec{r}^{\prime}|^3}  d^3\bvec{r}, 
\label{biot_savart}
\end{equation}
where $\mu_0$ is the vacuum permeability. Since the current co-propagates in the two graphene sheets of TBG in the "chiral" phase, the magnetic fields induced by the supercurrents add constructively making this particular feature of the TRS breaking phase measurable in experiment.

\bibliography{moire_sc.bib,DK_ref.bib}